\documentclass[sigconf,natbib=true]{acmart}

\usepackage[inline]{enumitem}

\usepackage{booktabs}
\usepackage{acronym}
\usepackage{subfig}
\usepackage{booktabs}
\usepackage{multirow}

\usepackage{xcolor}

\definecolor{mygreen}{rgb}{0.0, 0.6, 0.0}

\newcommand{\mypar}[1]{\vspace{0.05in}\noindent \textbf{#1 \,}}

\acrodef{CF}{collaborative filtering}
\acrodef{CRS}{conversational recommender system}
\acrodef{RL}{reinforcement learning}
\acrodef{TDM}{tree-based deep model}
\acrodef{LR}{logistic regression}
\acrodef{CF}{collaborative filtering}
\acrodef{DIN}{deep interest network}
\acrodef{TDM}{tree-based deep model}
\acrodef{NCF}{neural collaborative filtering} 
\acrodef{FM}{factorization machine}
\acrodef{GBDT}{gradient boosting decision tree}
\acrodef{MAB}{multi-armed bandit}
\acrodef{RL}{reinforcement learning}
\acrodef{DQN}{deep Q-network}
\acrodef{DRL}{deep reinforcement learning}

\acrodef{MTG}{multi-target ground truth}
\acrodef{STG}{single-target ground truth}
\acrodef{MNAR}{missing not at random}
\acrodef{MAR}{missing at random}

\acrodef{PMF}{probabilistic matrix factorization}
\acrodef{ExpoMF}{exposure matrix factorization}

\newcommand{\myeqp}[1]{\hyperref[eq:#1]{Equation~\ref*{eq:#1}}}
\newcommand{\mysec}[1]{\hyperref[sec:#1]{Section~\ref*{sec:#1}}}
\newcommand{\mytable}[1]{\hyperref[tab:#1]{Table~\ref*{tab:#1}}}
\newcommand{\myfig}[1]{\hyperref[fig:#1]{Figure~\ref*{fig:#1}}}
\newcommand{\myappendix}[1]{\hyperref[appendix:#1]{Appendix~\ref*{appendix:#1}}}
\newcommand{\myalg}[1]{\hyperref[alg:#1]{Algorithm~\ref*{alg:#1}}}

\AtBeginDocument{%
  \providecommand\BibTeX{{%
    \normalfont B\kern-0.5em{\scshape i\kern-0.25em b}\kern-0.8em\TeX}}}

\copyrightyear{2022}
\acmYear{2022}
\setcopyright{acmcopyright}
\acmConference[CIKM '22] {Proceedings of the 31st ACM International Conference on Information and Knowledge Management}{October 17--21, 2022}{Atlanta, GA, USA.}
\acmBooktitle{Proceedings of the 31st ACM International Conference on Information and Knowledge Management (CIKM '22), October 17--21, 2022, Atlanta, GA, USA}
\acmPrice{15.00}
\acmISBN{978-1-4503-9236-5/22/10}
\acmDOI{10.1145/3511808.3557220}

\ccsdesc[500]{Information systems~Recommender systems}
\ccsdesc[500]{Information systems~Evaluation of retrieval results}

\begin{document}

\title{KuaiRec: A Fully-observed Dataset and Insights for Evaluating Recommender Systems}




\author{Chongming Gao}
\authornote{Both authors contributed equally to this research.}
\email{chongming.gao@gmail.com}
\orcid{0000-0002-5187-9196}
\affiliation{%
  \institution{University of Science and Technology of China}
  \country{}
}

\author{Shijun Li}
\authornotemark[1]
\email{lishijun@mail.ustc.edu.cn}
\affiliation{%
  \institution{University of Science and Technology of China}
  \country{}
}

\author{Wenqiang Lei}
\authornote{Corresponding author.}
\email{wenqianglei@gmail.com}
\affiliation{%
  \institution{Sichuan University}
  \country{China}
}

\author{Jiawei Chen}
\email{sleepyhunt@zju.edu.cn}
\orcid{0000-0002-4752-2629}
\affiliation{%
  \institution{Zhejiang University}
  \country{}
}

\author{Biao Li}
\email{libiao@kuaishou.com}
\affiliation{%
  \institution{Kuaishou Technology Co., Ltd.}
  \country{}
}

\author{Peng Jiang}
\email{jiangpeng@kuaishou.com}
\affiliation{%
  \institution{Kuaishou Technology Co., Ltd.}
  \country{}
}

\author{Xiangnan He}
\email{xiangnanhe@gmail.com}
\orcid{0000-0001-8472-7992}
\affiliation{%
  \institution{University of Science and Technology of China}
  \country{}
}

\author{Jiaxin Mao}
\email{maojiaxin@gmail.com}
\affiliation{%
  \institution{Renmin University of China}
  \country{}
}

\author{Tat-Seng Chua}
\email{chuats@comp.nus.edu.sg}
\orcid{0000-0001-6097-7807}
\affiliation{%
  \institution{National University of Singapore}
  \country{}
}

\renewcommand{\shortauthors}{Gao et al.}

\begin{abstract}

The progress of recommender systems is hampered mainly by evaluation as it requires real-time interactions between humans and systems, which is too laborious and expensive. This issue is usually approached by utilizing the interaction history to conduct offline evaluation. However, existing datasets of user-item interactions are partially observed, leaving it unclear how and to what extent the missing interactions will influence the evaluation.
To answer this question, we collect a fully-observed dataset from Kuaishou's online environment, where almost all $1,411$ users have been exposed to all $3,327$ items. To the best of our knowledge, this is the first real-world fully-observed data with millions of user-item interactions.

With this unique dataset, we conduct a preliminary analysis of how the two factors — data density and exposure bias — affect the evaluation results of multi-round conversational recommendation. Our main discoveries are that the performance ranking of different methods varies with the two factors, and this effect can only be alleviated in certain cases by estimating missing interactions for user simulation. This demonstrates the necessity of the fully-observed dataset.
We release the dataset and the pipeline implementation for evaluation
at \textcolor{magenta}{\url{https://kuairec.com}}.

\end{abstract}

\keywords{Fully-observed data; Recommendation; Evaluation; User simulation}

\maketitle

\section{Introduction}\label{sec:intro}

Recommender systems (RSs) are designed to estimate users' preferences on items. A thorny problem in recommendation is how to faithfully evaluate the model, i.e., to tell whether a model can predict the user preference correctly. The most straightforward solution is to conduct the A/B test, which resorts to the online environment to see if real users are satisfied with the recommendations made by the model. However, this is time- and money-consuming and entails the risk of failure \cite{li2015toward,jagerman-when-2019}. Therefore, it is necessary to evaluate on the offline observed data. 

\begin{figure}[!t]
  \tabcolsep=0pt
  \centering
  \includegraphics[width=1\linewidth]{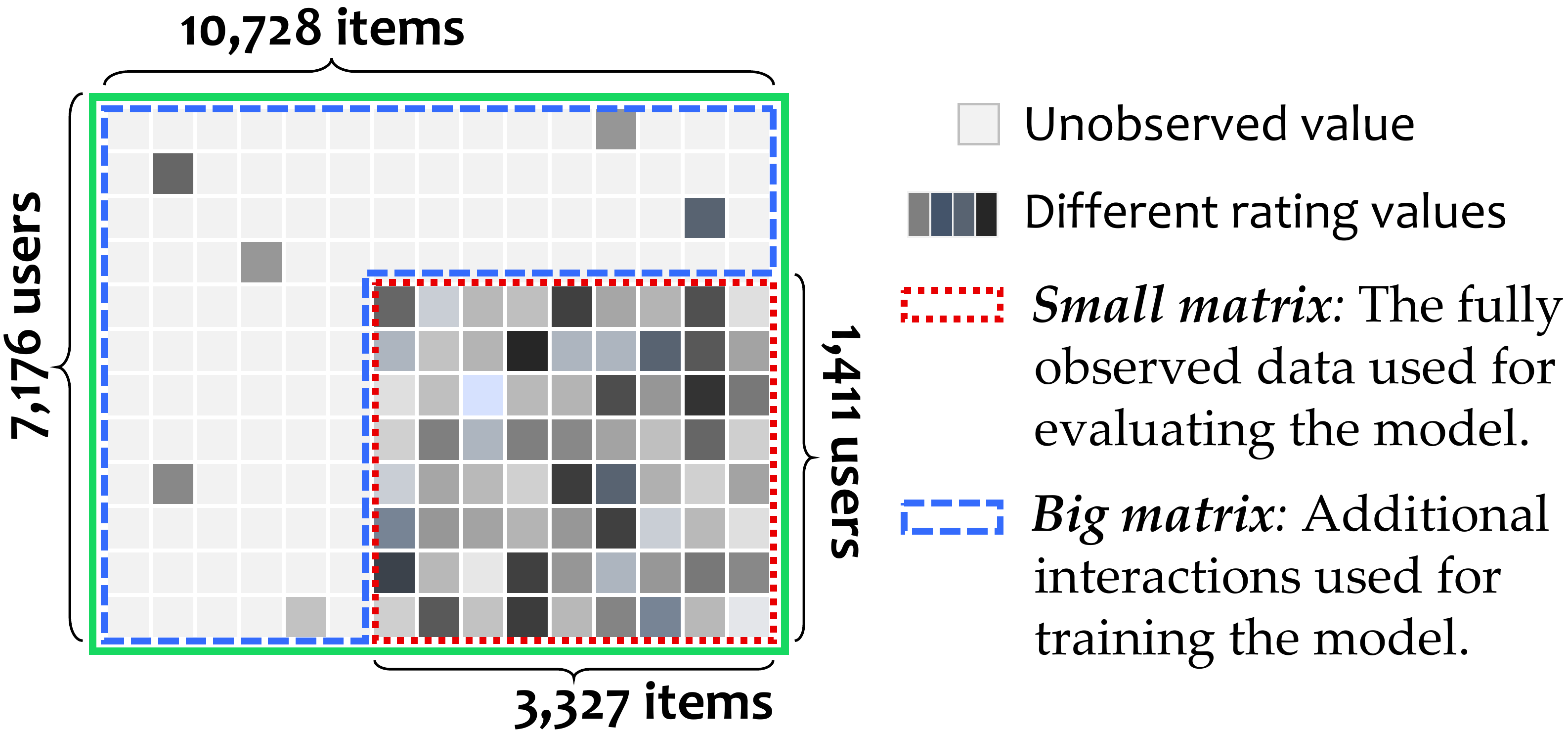}
   \vspace{-3mm}
  \caption{Illustration of the proposed \textcolor{mygreen}{\emph{KuaiRec}} dataset.}
   \vspace{-3mm}
  \label{fig:KuaiRec}
\end{figure}

However, the observed data in almost all recommendation datasets is highly sparse, i.e., only a few values are known in the user-item interaction matrix. This imposes great difficulties on evaluation. Specifically, the massive amount of missing values cannot be simply treated as either negative samples or missing-not-at-random (MNAR) data \cite{little2019statistical}. It means that the sparsely-observed data is not randomly sampled from the whole user-item matrix, which results in the exposure bias permeating through the data generation process in recommendation. The exposure bias can be further divided according to the cause. For example, \emph{popularity bias} occurs when the model is prone to exposing the popular items \cite{abdollahpouri2020multi}; 
And \emph{positivity bias} (or selection bias) is introduced because users have the propensity to select more often the items they like \cite{10.1145/2365952.2365982,huang-2020-keeping}. With these unknown values and pervasive biases in the recommendation data, the results of the offline evaluation are hard to be persuasive, which creates the widely-recognized challenge for evaluating recommender systems \cite{Marlin2009Collaborative,de2014reducing}.

To remedy the inherent defect of the data, some researchers present recommendation datasets containing randomly sampled data. For example, both the Yahoo! \cite{Marlin2009Collaborative} and Coat \cite{rs_treatment} datasets contain a set of missing-complete-at-random (MCAR) data that can be used for unbiased evaluation. However, the MCAR data of these datasets is still too sparse, which can bring a high variance to the evaluation results \cite{saito2021open}.


\subsection{A Fully-observed Dataset}

In this paper, we try to fundamentally address this issue by introducing \emph{KuaiRec}, a fully-observed dataset collected from the social video-sharing mobile App, Kuaishou\footnote{\url{https://www.kuaishou.com/cn}}. The term ``fully-observed'' means there are almost no missing values in the user-item matrix, i.e., each user has viewed each video and left feedback. This fully-observed dataset, dubbed the \emph{small matrix} for convenience, can be used for faithful evaluation. And for the training purpose, we further collect a larger dataset, dubbed the \emph{big matrix}, which contains additional interactions for the users and items in the \emph{small matrix}. 
\myfig{KuaiRec} gives an illustration of the small matrix and big matrix that make up the KuaiRec dataset.
Note that all user-item interactions in the \emph{small matrix} are excluded from the \emph{big matrix} to separate the training and evaluation
data. \mytable{dataset} lists the statistics of KuaiRec. Besides the user-item interaction, we further collect the side information of users and items. On the user side, we collected $30$ fields representing users' personal features, which includes $18$ encrypted one-hot vectors describing users' demographic information (such as age, gender, and city) as well as their personal behaviors. We also collect the social relationship among these users. On the item side, we list each item's tags, i.e., a set of categories such as \{Gaming, Sports\}. Furthermore, we collect $56$ fields to enhance the detailed information of items, which includes $45$ fields describing the statistics of each item within each day from July, 5th, 2020 to September 5th, 2020. For example, how many users click, like, or add a video to their favorite list each day.

To the best of our knowledge, KuaiRec is the first dataset generated from real-world recommendation logs with all users’ preferences on items known, and its scale (millions of user-item interactions) is much larger than the existing MCAR datasets. This can make the evaluation in offline data as effective as the online A/B test since there is no need to handle the missing values, which can benefit many research directions such as:
\begin{itemize}[leftmargin=*]
    \item \textbf{Unbiased recommendation}, which aims to explicitly identify the biases from data and remove the bad propensity from the recommender \cite{rs_treatment,saito_unbiased,chen2020bias}.
    \item \textbf{Interactive recommendation}, which focuses on improving the recommendation policy according to user feedback in the interaction on the fly \cite{Pseudo-Dyna-Q,zhang2019reward,gao2022cirs,wang2022best}.
    \item \textbf{Conversational recommender system (CRS)}, which is a kind of interactive recommender system that further utilizes the abundant features of items (e.g., tags, natural language) to efficiently and flexibly identify user preferences \cite{gao2021advances,lei2020interactive,li2020seamlessly,tutorial_CRS}.
\end{itemize}

\begin{table}[!t]
\tabcolsep=6.2pt
\caption{Statistics of the KuaiRec dataset. The density of small matrix is 99.6\% instead of 100\% because some users have blocked the videos from certain authors. All friends in the social network are in the 7,176 users of the big matrix.
}
\label{tab:dataset}
\vspace{-2mm}
\begin{tabular}{@{}lcccc@{}}
\toprule
                      & \textbf{\#users} & \textbf{\#Items} & \textbf{\#Interactions} & \textbf{Density} \\ \midrule
\textit{Small matrix} & 1,411            & 3,327            & 4,676,570               & 99.6\%        \\
\textit{Big matrix}   & 7,176            & 10,728           & 12,530,806              & 16.3\%           \\ \bottomrule
\end{tabular}
\tabcolsep=7pt
\begin{tabular}{@{}ll@{}}
\toprule
\textbf{User feature}:  & \begin{tabular}[c]{@{}l@{}} Each user has $30$ features which includes $12$ \\ explicit features and $18$ encrypted vectors. \end{tabular}                                                                                   \\ \midrule
\textbf{Item feature}:  & \begin{tabular}[c]{@{}l@{}} Each video has at least 1 and at most 4 tags\\  out of the totally 31 tags, e.g., \{Sports\}. \\ Each item has 56 explicit features, where \\ 45 fields are the statistics of each day. \end{tabular}                                                                                   \\ \midrule
\textbf{Social network}: & \begin{tabular}[c]{@{}l@{}}\textit{Small matrix}: 146 users have friends.\\ \textit{Big matrix}: 472 users have friends.\end{tabular} \\ \bottomrule
\end{tabular}
\vspace{-2mm}
\end{table}


\subsection{Insights in Evaluating Recommendations}
To demonstrate the efficacy and advantage of KuaiRec, we leverage it to conduct the evaluation of the multi-round conversational recommendation \cite{lei20estimation,10.1145/3404835.3462913,xu2021adapting}, a simplified setting in CRSs, which has caught the eyes of the research community recently due to its great potential \cite{gao2021advances}. 
Existing solutions for the evaluation of CRSs usually use the user simulation techniques based on the sparse partially-observed data. However, the trustworthiness of such approaches remains unknown \cite{gao2021advances,simulation2020kdd}. With KuaiRec, we can explore the research question that cannot be answered in existing studies:
\textit{Is partially-observed data as trustworthy as the fully-observed data w.r.t. the evaluation of CRSs?}

As the first attempt to explore the question, we examine two essential factors in partially-observed data: 
data density and exposure bias. 
The \textbf{data density} means the ratio of the items exposed to the users, i.e., the proportion of observed values in the user-item matrix. 
For the \textbf{exposure bias}, we explore three exposure strategies, namely \emph{uniformly random exposure} (which is unbiased), 
\emph{positivity-oriented exposure} (making exposed user-item interactions contain more historically positive samples to simulate the positivity bias \cite{abdollahpouri2020multi}),
and \emph{popularity-oriented exposure} (making the exposure lean towards the popular samples to simulate the popularity bias \cite{10.1145/2365952.2365982,huang-2020-keeping}).
We can examine the effect of the two factors by sampling part of the user-item interactions from the small matrix as the test set to conduct evaluations.
The experimental results on KuaiRec provide three insights:
1) Exposure biases caused by the exposure strategy greatly affect the performances and rankings of different models in evaluation. 
2) Even under the uniformly random exposure strategy (i.e., without exposure biases), different data densities can still result in inconsistent results.
3) The remedy that conducts matrix completion on the missing values can only partially alleviate the problem.
These insights entail the significance of our fully-observed dataset.




\noindent
We summarize our contributions as follows:
\begin{itemize}[leftmargin=*]
\item We are the first to present a real-world fully-observed dataset (density: almost 100\%) in recommendation. It contains millions of dense interactions and rich side information.
\item With this unique dataset, we design experiments to illustrate how data density and exposure bias affect the evaluation of recommendations. We further study the effect of estimating the missing values, i.e., matrix completion, on the evaluation results.
\end{itemize}

\noindent
We release this dataset as well as the scripts for loading and analyzing its statistics at \textcolor{magenta}{\url{https://kuairec.com}}. We also release the pipeline implement for evaluating conversational recommendation at \textcolor{magenta}{\url{https://github.com/xiwenchao/fully_observed_demo}}, with the hope to illustrate the advantage of this dataset and support further discussions along with related research topics. 

\section{Related Work}
In this section, we briefly review the datasets and problems in offline evaluation of recommender systems. Then, we introduce conversational recommender systems.

\subsection{Offline Evaluation in Recommendation} 

Online A/B tests have become ubiquitous in tech companies for the purpose of assessing the performances of models and rolling out the improved recommender system \cite{gilotte2018offline}. However, it usually consumes much time and money and thus is impractical for academic researchers to conduct the evaluation online.

Therefore, researchers usually resort to offline computing indicators in offline data, e.g., Precision, Recall, NDCG \cite{NDCG}, and MAP \cite{MAP}. However, such evaluations suffer from strong assumptions, such as independence between items and user feedback can be translated into a supervised task \cite{gilotte2018offline,marlin2012collaborative}. This is inconsistent with the nature of the recommender system, which should be treated as a sequential decision problem \cite{S3Rec_zhou,xinxin20,10.1145/3404835.3462855,saito2021open}. 

To address this problem, researchers propose two branches of solutions based on the offline data: 
1) off-policy evaluation (OPE), which aims to estimate the performance of the target policy using data generated by a different policy \cite{swaminathan2015counterfactual,lefortier2016large};
2) user simulation, whose core idea lies in filling the missing values before the evaluation \cite{simulation2020kdd,ie2019recsim}.
However, while the former solution suffers from the high variance issue \cite{saito2021open}, the latter will inevitably introduce estimation error. The primary cause of the problems is that the offline data is too sparse. In this way, fully-observed data is entailed for exploring how much the sparsity affects the results of the evaluation.

Up to now, far too little attention has been paid to collecting such a dataset. 
We briefly introduce classic existing high-quality datasets that contain unbiased randomly-sampled data used for offline evaluation in recommendation.
\begin{itemize}[leftmargin=*]
    \item \textbf{Yahoo!} \cite{Marlin2009Collaborative}. It contains the conventional missing-not-at-random (MNAR) data that contains approximately 300,000 user-supplied ratings from 15,400 users on 1,000 items in total. It also contains a set of missing-complete-at-random (MCAR) data by asking 5,400 users to give ratings on 10 items that are randomly selected from 1,000 items.
    \item \textbf{Coat} \cite{rs_treatment}. It collects the ratings of 290 users on 24 self-selected items and 16 randomly-selected items from total 300 items.
    \item \textbf{Open Bandit Dataset} \cite{saito2021open}. It contains interactions collected from two logged bandit policies: a Bernoulli Thompson Sampling policy and a uniform (random) policy. There are approximately 26 million interactions collected on users' clicks on 80 items.
\end{itemize}
However, these datasets are still highly sparse, e.g., Yahoo! has only 54,000 randomly-selected interactions out of $5,400 \times 1,000$ user-item pairs (i.e., density: $1\%$). So far, there is no dataset containing dense interaction data, not to mention the fully-observed data, to conduct the faithful evaluation in recommendation.

\subsection{Conversational Recommendation}
\label{sec:CR}
Recommender systems are powerful tools to help users reach their desirable items from a massive amount of items. However, the traditional static recommendation systems have limitations regarding capturing precious real-time user preferences \cite{zhangyang,wang2020click} and interpreting user motivation \cite{ma2019learning,gao2019bloma,yu2019generating}. Therefore, researchers overcome these problems by developing models that work in an online manner, i.e., interactive recommendation systems, such as the \ac{MAB} \cite{cikm13/ICF,christakopoulou2016towards,wang2018online,li2020seamlessly} and \ac{DRL} based models \cite{zhao2018recommendations,chen2019large,xian2019reinforcement,zheng2018drn,gao2022cirs}. 

\begin{figure}[!t]
  \tabcolsep=0pt
  \centering
  \includegraphics[width=.9\linewidth]{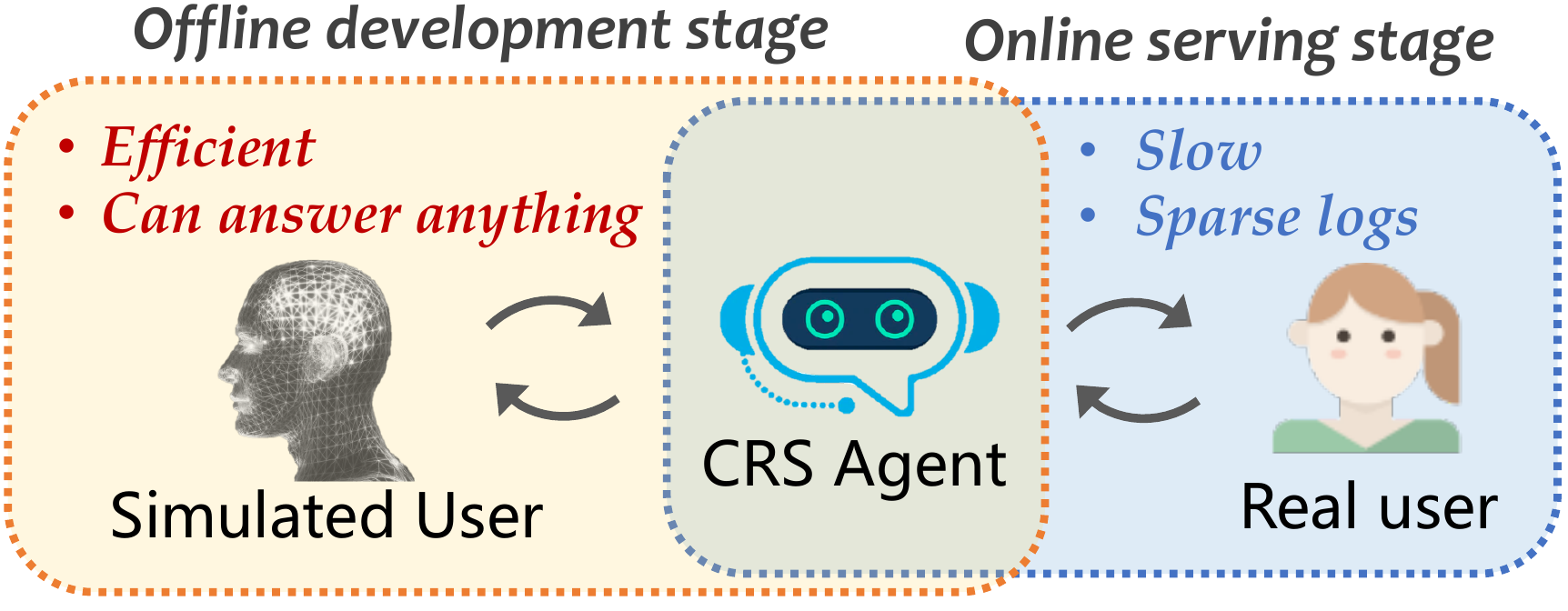}
  \caption{Interactions between the CRS and the simulated/real user on the model development/serving stage.}
  \vspace{-3mm}
  \label{fig:interaction}
\end{figure}

However, most interactive recommendation methods aim to collect user feedback on all recommended items, which suffer from low efficiency as there are too many of them. 
Critiquing-based methods, as an early form of \acp{CRS}, ask user questions about attributes to narrow down the item candidate space \cite{chen2012critiquing,luo2020deep,luo2020latent}. 
However, there is still a limitation: the models keep alternating asking and recommending, which should be avoided as the recommendation should only be made when the confidence is high. 
Recently, the emergence of \acp{CRS} has solved the problem. Compared with previous methods, a \ac{CRS} model has an additional user interface that allows for more flexible forms of interaction, e.g., in the form of tags \cite{christakopoulou2016towards,christakopoulou2018q}, template utterances \cite{lei20estimation,lei2020interactive,li2020seamlessly,10.1145/3404835.3462913,10.1145/3485447.3512088}, or free natural language \cite{10.1145/3394592,nips18/DeepConv,chen-etal-2019-towards,zhou2020improving,liu2020ACL,zhou2020topicguided,li2022user,ren2022variational,liu2021durecdial,zhou2022c2}. More importantly, a \ac{CRS} model has a conversation strategy module that controls the core logic of the multi-turn interaction \cite{lei20estimation,lei2020interactive,li2020seamlessly,zhang2020conversational,xu2021adapting, liu2020ACL, lewis-etal-2017-deal}.
Since our goal is to illustrate the necessity of the proposed fully-observed dataset in the offline evaluation, we focus on the multi-turn conversational recommendation which pays attention to the core interaction logic without the need to process the raw natural language.

Nonetheless, there are still a lot of challenges in \acp{CRS} needed to be addressed \cite{gao2021advances}.
One of the main challenges is the evaluation of \acp{CRS}.
To achieve that, researchers have to build user simulators to evaluate \acp{CRS} before serving online users (\myfig{interaction}). 
However, they are typically learned on the sparsely-observed data which usually contains various biases. \cite{gao2021advances, huang-2020-keeping} point out that the evaluation of \acp{CRS} by these biased user simulators might not be trustworthy. Thereby, the fully-observed dataset proposed in this work can serve as a necessary testbed for evaluating CRSs.

\section{Data Collection}
\label{sec:dataset}

In this section, we introduce the data collection process and show the representativeness of the collected data.
 
\subsection{The KuaiRec Dataset}
\label{sec:kuairecdata}
We collect the data from Kuaishou App, a famous short-video platform with billions of users. On this platform, users can watch a variety of short-form videos from genres such as dance, entertainment, and fitness/sports. 
The videos are organized by recommending streaming where each time the user can see only one video. The user can swipe up or down to skip to the last or next videos at any time without having to wait for the end of the video. In this paper, we sometimes refer to videos as \emph{items} for ease of understanding.

\subsubsection{Collecting Fully-observed Data (i.e., The Small Matrix)}
All the users and videos in our dataset, as well as their interaction records, are collected in the period from July, 5th, 2020 to September 5th, 2020 on Kuiashou App.
Over this period, we first collect a set of videos that are excluded from all advertisements (i.e., sponsored items), junk videos, and spammers. Then we further select a subset of users and videos from them, aiming to enable as many selected videos as possible to have been watched by the selected users (i.e., high density). For the missing values, i.e., the rest videos that the user did not watch, we manipulate the online recommendation rule to insert these videos into the recommendation streaming to make sure all of them have been shown to these users and received feedback. It takes $15$ days for this exposure process. 

Finally, there are $1,411$ users and $3,327$ videos that compose a matrix where each element represents a user's feedback on a video. The density of this matrix is $99.6\%$ instead of $100\%$ because some users have explicitly blocked the videos from certain authors. We refer to this fully-observed matrix as the \emph{small matrix}.
We admit that the high density can inevitably bring in the popularity bias \cite{abdollahpouri2020multi}, however, it can still serve as a trustworthy testbed to evaluate and compare different recommendation models.


\subsubsection{The Big Matrix: Peripheral Data of the Small Matrix}
The fully-observed dataset can be used for the trustworthy evaluation of \acp{CRS}. However, we need additional data for the training of \acp{CRS} \cite{Sun:2018:CRS:3209978.3210002,lei20estimation,lei2020interactive}. 
Hence, we collect a partially-observed data that contains the interactions of more users and videos.
We refer to this larger data as the \emph{big matrix}. It contains $7,176$ users and  $10,728$ videos that include all the users and videos in the \emph{small matrix}. 
Note that all user-item interactions in the \emph{small matrix} are excluded from the \emph{big matrix} to separate the training and evaluation data. 

\begin{figure}[!t]
  \tabcolsep=0pt
  \centering
  \includegraphics[width=0.98\linewidth]{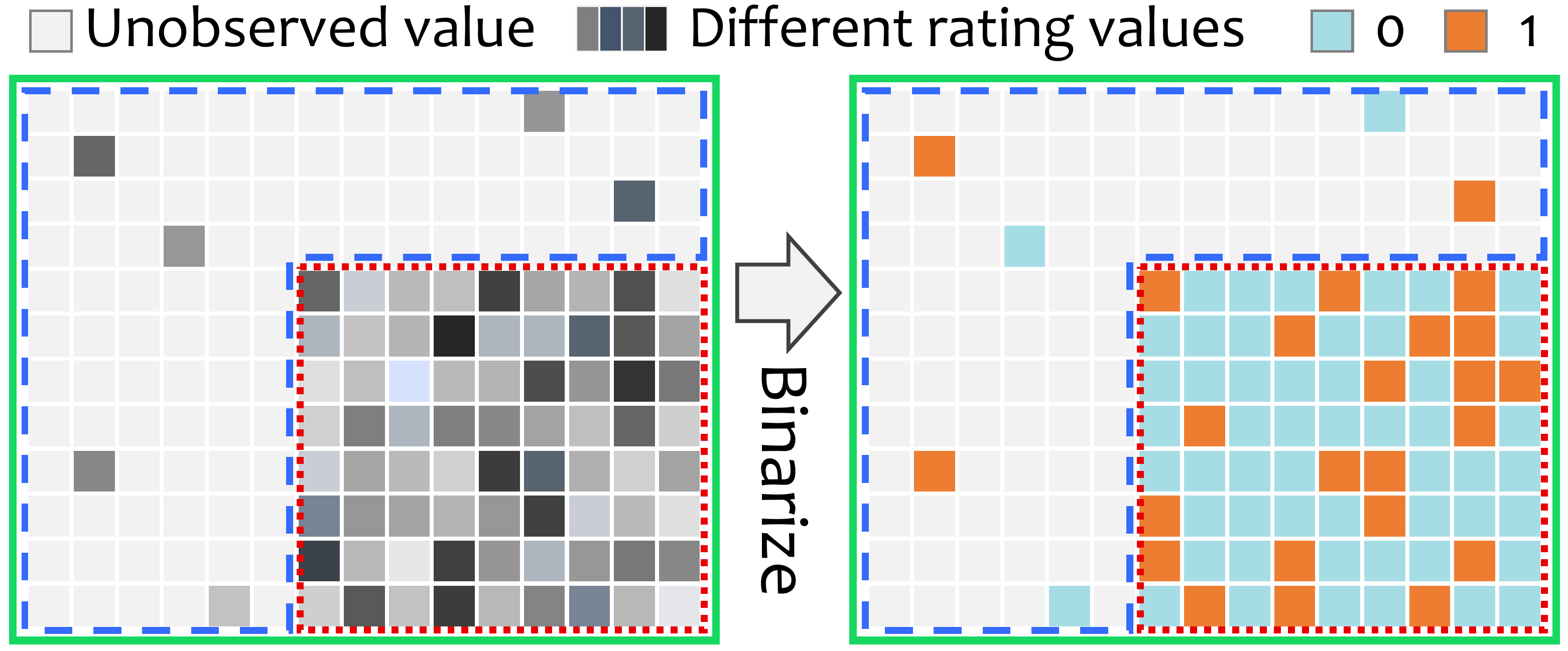}
  \caption{Transformation from the rating matrix to the binary matrix to define the positive samples.}
  \vspace{-3mm}
  \label{fig:binarize}
\end{figure}

\subsubsection{Defining the Positive Interactions}
Our collected raw data contains various user feedback on each watched video such as comments, likes, the total time spent on the video, etc. Without losing generality, we binarize each user-video pair into either positive or negative samples accordingly to the watch time. Specifically, we define a user-video pair as a positive sample if the user's cumulative watching time of this video is greater than twice the duration of this video, i.e., the user has watched this video completely at least twice. This is deemed as a strong indicator of the users' preference for the videos suggested by the data analysis team of Kuaishou. The process is illustrated in \myfig{binarize}. We did not use the raw ``Like'' signal left by users because it is too sparse.



\subsubsection{Side Information Collection}

We collect the rich features for the users and videos in the \emph{big matrix} as shown in \mytable{dataset}. On the user side, we collected $30$ fields representing users' personal features, which includes $18$ encrypted one-hot vectors. Besides, people's minds can be affected by their friends. Therefore, we collect the social network of the $7,176$ users in the \emph{big matrix} to benefit the social recommendation \cite{yu2018adaptive,yu2019generating}.

On the item side, we collect $56$ fields to enhance the detailed information of items, which includes $45$ fields describing the statistics of each item within each day from July, 5th, 2020 to September 5th, 2020. For example, how many users click, like, or add a video to their favorite lists each day. Especially, we collect $31$ genres (i.e., tags) of videos. These tags are manually defined and annotated by the internal annotation team of Kuaishou. Each video is related to at least one and at most four tags. These tags, such as fitness/sports and recipes, are very useful in the question asking and candidate reduction in conversational recommendation models.
Armed with these features, researchers can study more comprehensive topics using this dataset.


\subsection{Hypothesis Testing of Representativeness}
\label{HT}
Now, we show that the users and videos in the collected fully-observed \emph{small matrix} are representative of all users and videos in the whole Kuaishou platform. 
Specifically, we first select six key features (see \mytable{Hypothesis_test}) for users and videos, respectively. These features are thought to be the most representative to describe the properties of the users/videos by the data analysis team of Kuaishou.
Then, we conduct the discrete two-sample Kolmogorov–Smirnov (KS) tests \cite{marsaglia2003evaluating} to verify whether the users (or videos) in the \emph{small matrix} have the same distribution, w.r.t. the six features, as all users (or videos) in the one-month data of Kuaishou. 

Our null hypothesis is that the two distributions  have \emph{no significant difference} w.r.t. each feature, i.e., samples from the two data are drawn from the same distribution. We conduct the tests under a $5\%$ rejection level and report the corresponding \emph{p-values} in \mytable{Hypothesis_test}.

It's clear to see that the \emph{p-values} of all KS tests on each feature are above the rejection level of $5\%$. It means that we cannot reject the null hypothesis, i.e., users and videos in the fully-observed data are sufficiently \emph{representative} to reflect the general properties of the users and videos on the whole Kuaishou platform.

\begin{table}[!t]
\tabcolsep=4pt
\caption{The \emph{p-values} of two-sample KS tests. The selected features for users are \{\emph{age, location, register channel, operation system, 1st purchase channel, 2nd purchase channel}\}. For videos, the features are \{\emph{background music class, visible state, client setting, sound track class, brand class, platform class}\}. The six features are denoted by Feat1–Feat6. 
}
\vspace{-2mm}
\label{tab:Hypothesis_test}
\begin{tabular}{@{}lcccccc@{}}
\toprule
                      Feature & \textbf{Feat1} & \textbf{Feat2} & \textbf{Feat3} & \textbf{Feat4} & \textbf{Feat5} & \textbf{Feat6}\\ \midrule
User   &   92.89\%   & 88.27\%    &    6.03\%     & 53.44\% &   19.00\%   &  6.56\%      \\ 
Video  &   88.27\%   & 97.62\%    &   99.96\%     & 31.97\% &   79.74\%   & 99.96\%      \\
\bottomrule
\end{tabular}
\end{table}

\section{Environment Setting for CRS}
\label{sec:env}

To better illustrate how to utilize the proposed dataset in evaluating recommendations, we focus on the Multi-round Conversational Recommendation (MCR) scenario \cite{lei20estimation}. We will start with the introduction of MCR as well as how the user simulator is built. Then we introduce how we synthesize the biased partially-observed data to explore its effect on the evaluation.

\subsection{Multi-round Conversational Setting}
\label{sec:mcr}
The MCR setting is thought to be the most straightforward and realistic conversational recommendation setting in research so far and has been widely employed in many CRSs \cite{lei2020interactive, lei20estimation, xu2021adapting, zhang2018towards, li2020seamlessly}.


MCR is a typical form-based \cite{Jannach2020ASO} setting, which means users interact with the system in pre-defined manners instead of dynamically generated natural languages. Thereby we can focus on the core interaction logic rather than language understanding.
In MCR, a multi-round interaction process between a \ac{CRS} model and a user (usually a simulated one) is called one conversation (a.k.a. one conversation session). The system can choose to either ask a question or recommend items to the user at each conversation turn. 
If the current turn is to recommend, the \ac{CRS} model recommends top-$K$ items and expects a response as ``Yes'' or ``No''. Otherwise, the \ac{CRS} model asks a question about attributes such as ``Would you like videos with the fitness/sport tag?'' Then the system will receive binary feedback from the user as ``Yes'' or ``No''. If the response from the user about querying attributes or items is negative, the system removes the corresponding videos from the candidate pool. If the user gives positive feedback to the asked attribute, the agent 
removes all items that do not contain the preferred attribute from the candidate pool.
The conversation will continue until the user accepts the recommended items. 
To best satisfy the users, the objective of \acp{CRS} in the MCR setting is to make successful recommendations with the fewest conversation turns. 

As discussed in \mysec{CR}, researchers built user simulators to automate the evaluation of \acp{CRS}. To simulate a user, the simulator needs to give feedback on the items or attributes based on the user preferences estimated from the existing observed data \cite{lei2020interactive, lei20estimation, li2020seamlessly, zhang2018towards, zhang2020conversational}. However, it remains unknown how sparsity and biases in the partially-observed data affect the evaluation.

\subsection{Synthesizing Partially-observed Data}
\label{sec:synthesize}

Although we can conduct a trustworthy evaluation using the fully-observed data, we want to find out how partially-observed data affect the evaluation. For example, how does the density of the partial data affect the evaluation? And, how do the biases introduced by exposure strategies in the partial-observed data affect the performances and rankings of different models? Therefore, we employ three types of widely used exposure strategies to generate \emph{partially-exposed data} in the \emph{small matrix} to synthesize the partially-observed data for evaluation. 

\subsubsection{Uniformly Random Exposure}
Uniformly random is an ideal way to get the data without exposure biases. The collected data is treated to be a reasonable substitute for fully-observed data.
We uniformly sample the elements in the \emph{small matrix} without replacement for $9$ times, with the density setting to $\{10\%, 20\%, \ldots, 90\%\}$ accordingly. The $9$ sampled \ac{MAR} data are used in our experiments to explore how the data density affects the evaluation of \acp{CRS}.

\subsubsection{Positivity-oriented Exposure}
In reality, the partially-observed data is usually biased towards the items that users like, which can be explained for two reasons. First, users tend to choose the items they like to consume or rate, a.k.a., selection bias \cite{marlin2012collaborative}. Second, the recommendation result can be dominated by the highly-confident items provided by the model \cite{10.1145/3397271.3401083}. Therefore, we employ positivity-oriented exposure to simulate such biased partially-observed data. Specifically, for each user in the \emph{small matrix}, we sample without replacement a certain number of items with respect to the positivity distribution calculated on the \emph{big matrix},
e.g., if item A is liked by $6$ users and item B is liked by $2$ users in the \emph{big matrix}, the exposure probability of A should be three times of B. 
Similarly, with varying the density in $\{10\%, 20\%, \ldots, 90\%\}$, we get $9$ sampled \ac{MNAR} datasets containing the positivity bias.

\subsubsection{Popularity-oriented Exposure}
\label{sec:popularity_sampling}
The recommendation model can also bias toward the popular items. Therefore, it is common for the partially-observed data to contain popularity bias \cite{abdollahpouri2020multi}. 
To satisfy the long-tail distribution requirement\footnote{Real-world recommendation data always satisfies a long-tail distribution \cite{10.1145/1454008.1454012,10.14778/2311906.2311916}}, 
we employ Zipf's law \cite{zipflaw} to decide the probability of each item to be exposed.
Specifically, for each user in the \emph{small matrix}, we set the probability of exposing the $k$-th popular item as $P(k)=\frac{1 / k^{s}}{\sum_{n=1}^{N}\left(1 / n^{s}\right)}$, where $s = 0.5$, $N=3,327$ (the number of items in the \emph{small matrix}) in our paper. This order of items is set as the order of popularity of items (the number of occurrences) in the \emph{big matrix}.
This leads the data to follow the long-tail distribution, while the original order of the popularity for items is preserved.
Again, we derive the $9$ sampled \ac{MNAR} datasets containing popularity bias with the data density equal to $\{10\%, 20\%, \ldots, 90\%\}$.

\section{Experiments}
\label{sec:experiments}
In this section, we conduct intensive experiments on the evaluation of \acp{CRS} on the KuaiRec dataset. 
\subsection{Experimental Setting}
\label{sec:experimental_setting}
\subsubsection{Research Questions}
we will organize our experiments and analyze the results based on the two questions:
\newline\textbf{RQ1:} How does the partially-exposed data (with and without biases) affect the evaluation of \acp{CRS}?
\newline\textbf{RQ2:} Can we improve the evaluation on the partially-exposed data by estimating the missing values (i.e., matrix completion)?

\subsubsection{Evaluation Setting}\label{sec:evaluation_setting}
We conduct experiments on the MCR setting, which has been introduced in \mysec{mcr} in detail.
We follow the configurations in EAR \cite{lei20estimation}.
Typically, we set the maximum round of each conversation to \emph{15}, since users always have limited patience. And the length of the recommendation list in each round is set to \emph{10}.
To explore the helpfulness of estimating the missing values, we compare the evaluation results of \acp{CRS} on different partially observed data before and after matrix completion.

\subsubsection{User Simulation}\label{sec:user-simu}

As discussed in \mysec{CR}, we need a user simulator to interact with the CRS. In our experiments, we simulate users from the partially-exposed matrix created on the \emph{small matrix}. 

It is worth mentioning that most existing studies \cite{Sun:2018:CRS:3209978.3210002,lei20estimation,lei2020interactive} used the \ac{STG} setting to evaluate \acp{CRS}. In the \ac{STG} setting, the target of each conversation is to recommend \emph{one} ground-truth item.
If the recommended top-$K$ items contain the ground-truth item, the user simulator accepts the recommendation and ends the conversation successfully. 
When the \ac{CRS} model asks a question, the user simulator gives positive feedback only if the asked attribute belongs to the ground-truth item. 

However, in real practice, users tend to have multiple preferences and diverse interests \cite{aziz2020strategyproof, lytras2021information}.
For example, in the video recommendation streaming, the user may like and accept more than one video recommended by the system. In this case, many videos liked by the user will be mistakenly treated as negative samples under the \ac{STG} setting. Therefore, we pay more attention to the \ac{MTG} setting where users have multiple ground-truth items in a session. Specifically, the user simulator accepts the recommendation only if \emph{one of} the ground-truth items is in the recommended top-$K$ items. When the \ac{CRS} model asks a question, the user simulator gives positive feedback if \emph{at least one} of the ground-truth items contains the queried attribute.

In the evaluation, the number of conversations equals the number of positive user-item pairs in the test set, i.e., we will conduct one conversation for each positive sample in both the \ac{STG} and \ac{MTG} settings. The difference is that in the \ac{MTG} setting we will  remove the accepted item from the ground-truth set after a successful recommendation for each simulated user. 

\subsubsection{Metrics}\label{sec:metrics}
Following~\cite{lei2020interactive,lei20estimation,Sun:2018:CRS:3209978.3210002}, we use two metrics to assess the performance of \ac{CRS} models in the multi-round conversation. The first is the Average Turns (AT) of all conversations. It is expected to be as small as possible because the model should make successful recommendations with the fewest rounds. Another metric is Success Rate (SR@$t$): the proportion of the conversations ended before (or at) round $t$ in all the conversations.

\subsubsection{Baselines}\label{sec:baselines}
We select four representative \ac{CRS} as baselines. All of them have a recommendation model to measure the similarity between users and items, as well as a conversation strategy to decide whether to ask attributes or make recommendations. We use the \ac{FM} model \cite{rendle2010factorization} as the recommendation engine in all four methods for ease of comparison. 
\begin{itemize}
	\item \emph{Max Entropy} \cite{Sun:2018:CRS:3209978.3210002}. 
	When asking questions, it always chooses an attribute with the maximum entropy within the current candidate item set. 
	\item \emph{Abs Greedy} \cite{christakopoulou2016towards}. It focuses on recommending items in every round without asking any questions and keeps updating the model by treating the rejected items as negative examples. 
	\item \emph{CRM} \cite{Sun:2018:CRS:3209978.3210002}. It records user preference in a belief tracker and uses \ac{RL} to find the optimal policy.
	\item \emph{EAR \cite{lei20estimation}} A classic \ac{CRS} model that contains a strategy module based on the \ac{RL} model similar to CRM, except it considers a more sophisticated state in \ac{RL}.
\end{itemize}

Since all the above models contain relatively complex components, i.e., the FM model, 
we further add four heuristic methods, each of which only contains a naive conversation strategy. 
\begin{itemize}
	\item \emph{Random}. It randomly selects $10$ items from the candidate set to recommend in every turn without asking any questions. 
	\item \emph{Popularity-oriented Recommender (PopRec)}. Recommending the top $10$ popular items in the candidate sets at every turn without asking any questions. 
	\item \emph{Positivity-oriented Recommender (PosRec)}. Similar to \emph{PopRec}, except it recommends the most positive items in the rest of the item candidates. The most positive items are the items loved by as many users as possible in \emph{big matrix}.
	\item \emph{Attribute-asking Recommender (AttrAskRec)}.
	It alternates asking attributes and recommending items. When asking a question, it asks about the most popular attribute; when recommending, it randomly recommends 10 items.
\end{itemize}

\subsection{Exploring the Effects of Partially-Exposed Data in CRS Evaluation}

\begin{figure}[!t]
  \tabcolsep=0pt
  \centering
  \includegraphics[width=.98\linewidth]{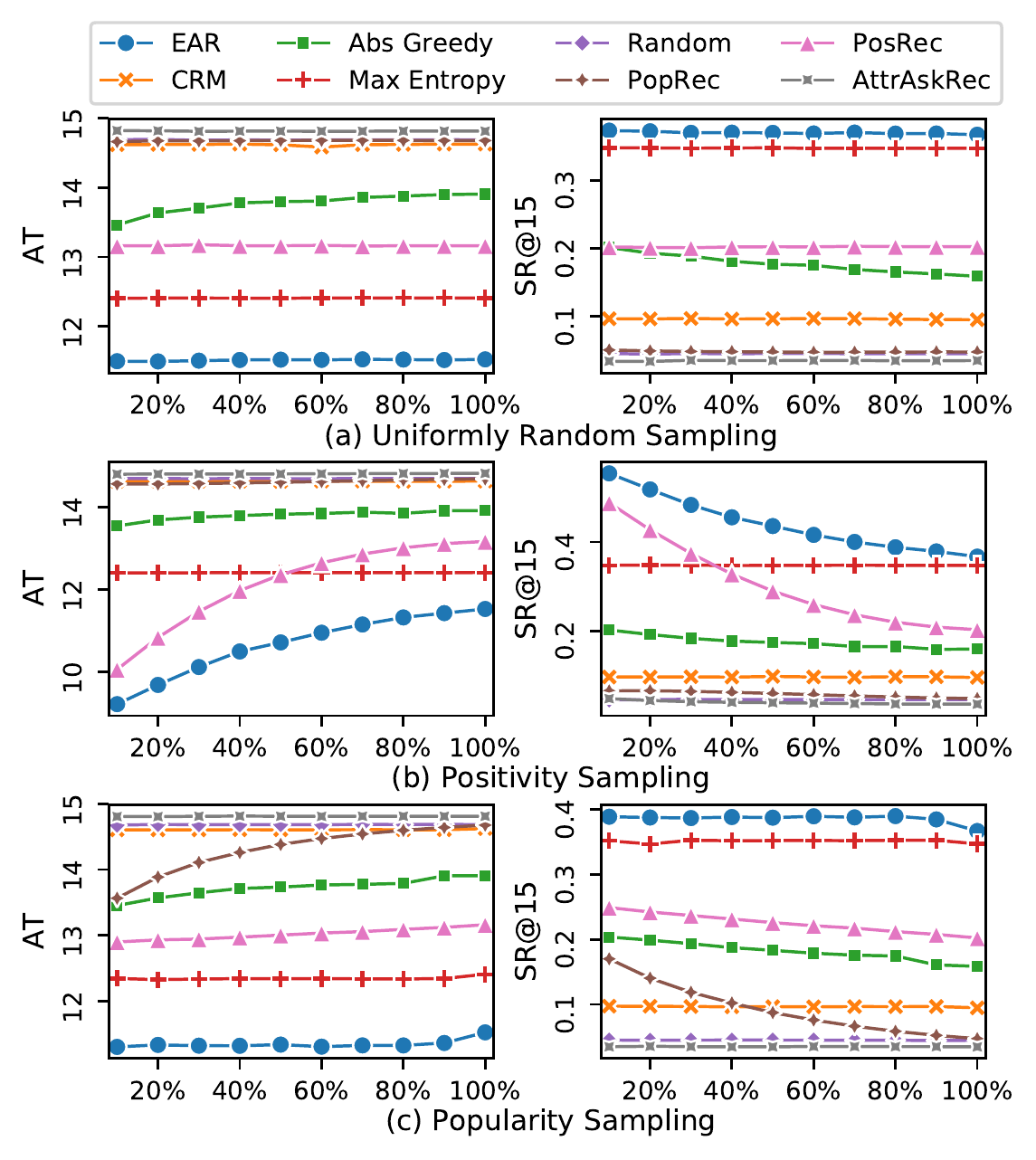}
  \caption{Performance of eight methods on the STG setting by varying the data density and exposure strategies. 
  }
  \label{fig:STG}
   \vspace{-2mm}
\end{figure}

We use the sampled partially-exposed data to explore how the density and exposure strategy affect the evaluation of \acp{CRS}. 
Specifically, we repeat the evaluation $10$ times and report the average AT and average SR@$15$ over all conversations. 
The partially-exposed data are generated from the \emph{small matrix} following the rules in \mysec{synthesize}.
It should be noted that all \ac{CRS} methods are already trained on the \emph{big matrix} before the evaluation.

Instead of paying attention to the absolute performance of the eight methods, we focus on 
how the rankings (i.e., the relative order) of these \acp{CRS} change w.r.t. various densities and exposure strategies. 
We start with the \ac{STG} setting and then make the counterpart experiments in the \ac{MTG} setting with the same configurations. 


\begin{figure}[t]
  \tabcolsep=0pt
  \centering
  \includegraphics[width=0.98\linewidth]{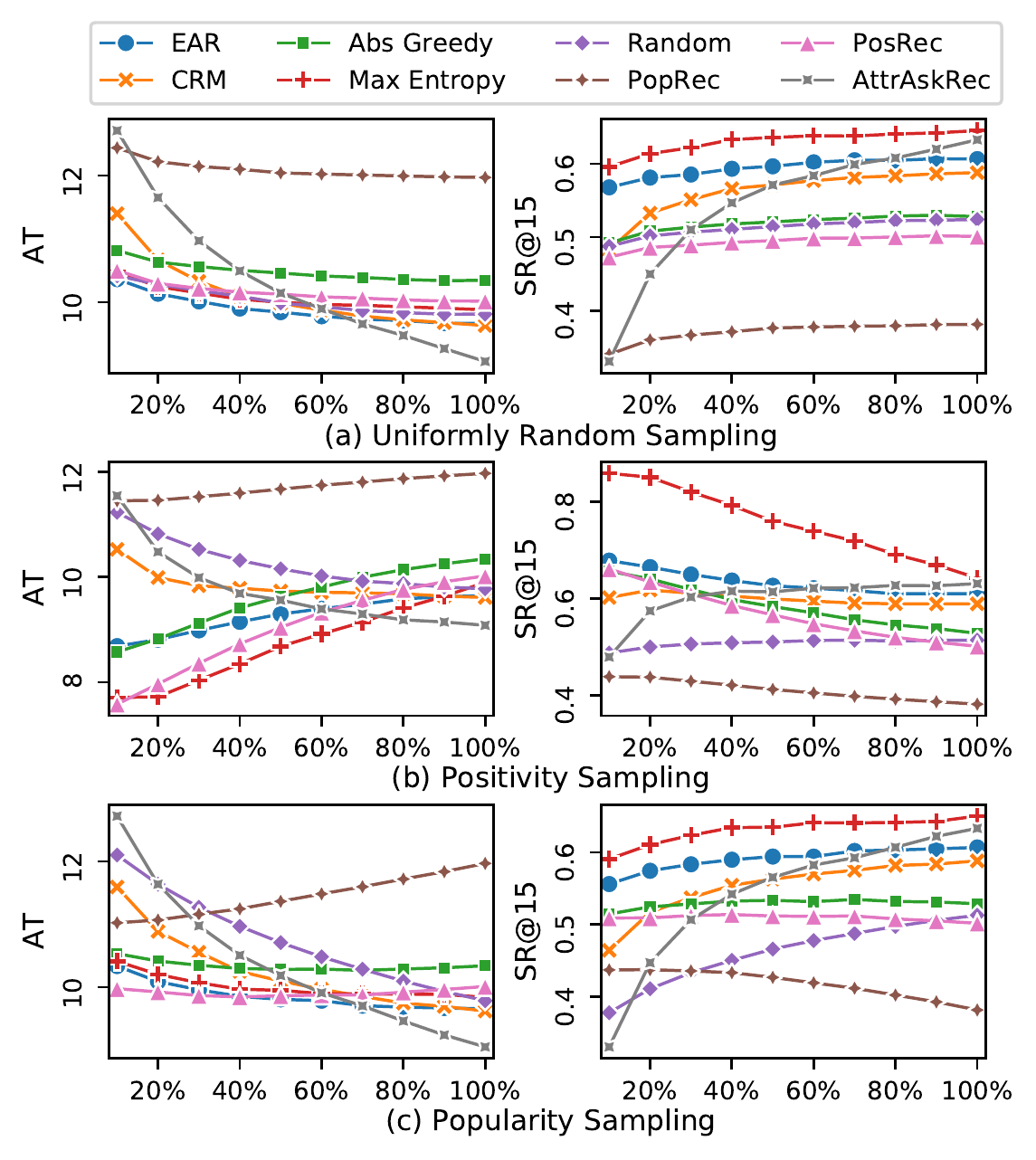}
  \caption {Performance of eight methods on the MTG setting by varying the data density and exposure strategy.}
  \label{fig:MTG}
  \vspace{-2mm}
\end{figure}

\subsubsection{Analyzing the Results under the STG Setting}
We illustrate the results of eight methods on the \ac{STG} setting under three exposure strategies and $10$ densities (including the fully observed data) in \myfig{STG}. From the results, we can get the following insights:

\mypar{Uniformly Random Exposed Data in Evaluation.}
In the results of uniformly random exposure, the performance rankings of eight methods keep unchanged and the performances almost keep stable (except for the \emph{Abs Greedy}) over all densities. This is intuitive as the positive samples under the uniformly random exposure are independent and identically distributed.
Thus the average performance will not vary much with different data densities due to the law of large numbers \cite{dekking2005modern}.


\mypar{Effects of Positivity Biases in Evaluation.}
In the results of positivity-oriented exposure, \emph{PosRec} performs satisfyingly at the beginning and becomes worse as the density increases. This is due to its recommendation mechanism (i.e., always recommending the most positive items in the \emph{big matrix}) is correlated with the mechanism of the exposure strategy. Specifically, when the data density is low, the exposed positive samples are the most positive items, therefore the \ac{CRS} can find each of them easily in a few turns, hence the AT is low and SR@15 is high. However, when the density increases, more items with a relatively low positivity ratio will be sampled in our synthesized testing dataset. Those items have low rankings in the recommended list of \emph{PosRec}. Therefore the performance deteriorates because of these ``hard to predict'' items. It is interesting that \emph{EAR} shows a similar trend, which demonstrates that \emph{EAR} also tends to recommend positive items.
Aside from \emph{EAR} and \emph{PosRec}, other methods have relatively stable absolute performance and consistent rankings, indicating that these methods are not vulnerable to the positivity bias under the STG setting.


\mypar{Effects of Popularity Biases in Evaluation.}
The results show a similar phenomenon: only \emph{PopRec} has unstable performance and inconsistent rankings. It is because \emph{PopRec} always recommends the most popular items, making it easy to find the correct items when the exposed items are the most popular ones. When the unpopular items are exposed with the data density increasing, the average performance of \emph{PopRec} is held back by the unpopular items which are hard to predict.
Other methods have stable performance and consistent rankings, indicating that they are not sensitive to the popularity bias under the STG setting.

\vspace{2mm} \noindent
In the \ac{STG} setting, each item is evaluated independently, which is straightforward for the learning and evaluation of \acp{CRS}. However, the users in most real-world applications have multi interests and there is more than one correct answer in the evaluation. Therefore, we explore how things work in the \ac{MTG} setting. 

\subsubsection{Analyzing the Results under the MTG Setting}
In the \ac{MTG} setting, we implement the counterpart experiments above and show the results in \myfig{MTG}.
It's obvious that the results are quite different compared to the results in the \ac{STG} setting. Even under the uniformly random exposure, both the absolute performance and rankings of the CRSs vary a lot on different testing data. 

\mypar{Challenges and Opportunities under the MTG Setting.}
The instability of the results comes from the complexity of the \ac{MTG} setting. As discussed in \mysec{user-simu}, a successful recommendation is achieved when at least one of the ground-truth items is contained in the recommended list. However, multiple ground-truth items do not mean that the recommendation task becomes easier. Instead, a system needs to estimate user preferences based on diverse feedback (recall that user simulators give feedback based on multiple preferred items). 
Under the \ac{MTG} setting, \emph{Max Entropy}, with a simple conversation strategy, surprisingly performs best on SR@15.
The reason that \emph{Max Entropy} outperforms \emph{EAR} might be that \emph{EAR} is specifically designed under the STG setting. However, there is no \ac{CRS} designing for the MTG setting until now. It remains to be a promising direction to develop the \ac{CRS} under the MTG setting since MTG is more realistic in real life.

\mypar{Effects of the Question Asking in the MTG Setting.} 
Since the \ac{MTG} setting is more complicated, the next question is how a model can perform well in this setting. 
From the results, we find some clues: we observe that asking attributes affects the performance in the MTG setting more significantly than in the STG setting. 

There are two empirical observations: First, the \emph{AttrAskRec} method performs worst in the STG setting, but it performs very well when the data density is large in the MTG setting. Second, \emph{CRM} outperforms \emph{Max Entropy} on AT at a certain point when the density increases. We found that \emph{CRM} tends to ask more questions than \emph{Max Entropy}. Specifically, we calculate the average probability for \emph{CRM} to ask a question in each turn is $81.7\%$, while \emph{Max Entropy} is $66.5\%$.

To understand the reason, we revisit the mechanism of the user simulator in the MTG setting. When being asked a question about an attribute, the user simulator will return positive feedback if one of the ground-truth items contains the queried attribute. Therefore, when there are few ground-truth items (when the density is small), many asked questions will be rejected, which wastes the interaction turn. Thus, the average turns will be high. Conversely, when there are more ground-truth items (when the data density is high), it will be easier for the CRS to get positive feedback after asking questions. The positive response can help the \ac{CRS} quickly narrow down the list of candidate items, which results in smaller average turns.

\mypar{Effects of Biases in the MTG Setting.}
Similar to the observation in the STG setting, the positivity biases and popularity biases can still affect the evaluation results of \acp{CRS}. 
More precisely speaking, the rankings of \acp{CRS} change more sharply.
For instance, \emph{Random} behaves badly when the density is low on data exposed with positivity bias. In this case, almost all exposed items have a high positivity ratio, so randomly recommending items can be a poor strategy. When the density increases and more items with a relatively low positivity ratio are exposed, the relative performance of \emph{Random} increases.
By contrast, the absolute performance of \emph{PosRec} decreases as the density increases. \emph{PosRec} only recommends items with maximal positivity ratio, hence it is unable to recommend items with low positivity ratio. Nevertheless, when the data is sampled uniformly randomly, the variation of absolute performance of both \emph{Random} and \emph{PosRec} is much milder. Therefore, the bias brought by the exposure strategy can aggravate the discordance of rankings.


 \subsection{Effects of Estimating the Missing Values}
We further investigate the effects of the remedy: estimating the missing values in the partially-exposed matrix and using the estimated positive samples for building the user simulator in evaluation. 

\subsubsection{Matrix Completion}
Estimating missing values, i.e., matrix completion is a well-studied research topic. Due to the limitation of space, we only select three methods related to our research: \ac{PMF} \cite{mnih2007probabilistic}, \ac{ExpoMF} \cite{expomf} and its variant ExpoMF-cov \cite{expomf}. All of these methods are classic baselines of the probabilistic methods in matrix factorization (MF). PMF is an ordinary MF method without considering biases in the data, while ExpoMF and ExpoMF-cov are more sophisticated methods since they use a random variable to model whether a user-item interaction happens. i.e., they can differentiate the unobserved event from a negative preference, which shows an advantage in situations where there are few exposed interactions. 
The difference between ExpoMF-cov and ExpoMF is that ExpoMF-cov adds the additional exposure covariates to take into account the additional item features. 

\begin{figure}[!t]
  \tabcolsep=0pt
  \centering
  \includegraphics[width=.95\linewidth]{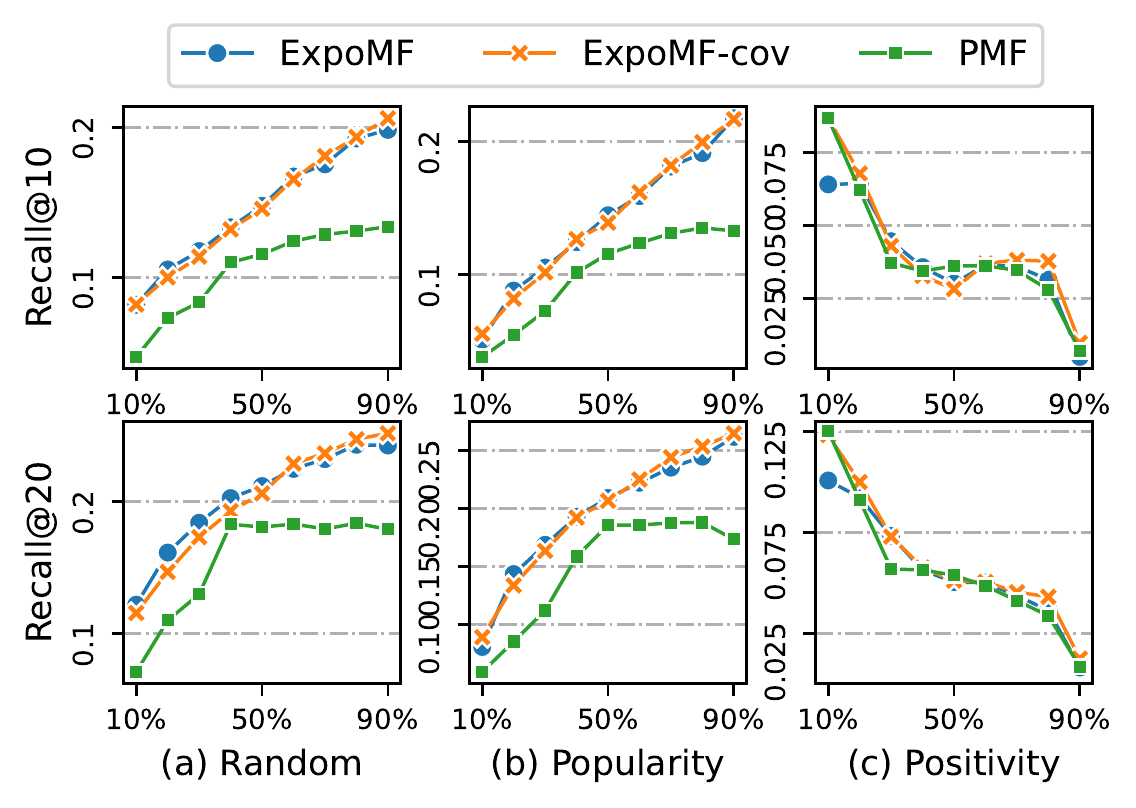}
   \vspace{-3mm}
  \caption {Performance of the matrix completion on the partially-exposed data. The training set is the exposed data, i.e., the observed data, and the test set is the unobserved data.}
  \vspace{-3mm}
  \label{fig:MF_performance}
\end{figure}

We train these methods on the observed part of all datasets with the density and exposure strategy varied and test the performance on the unobserved part, i.e., unsampled values. We report the Recall@$10$ and Recall@$20$ on all datasets and illustrate the results in \myfig{MF_performance}. From the results, we observe the following insights:

\mypar{Effects of Explicit Modeling the Exposure Event.}
PMF shows an inferior performance to two debiased methods (ExpoMF and ExpoMF-cov), especially under uniformly random exposure and popularity-oriented exposure. It is due to the fact that ExpoMF and ExpoMF-cov take into account whether the value $0$ in the training data represents an unobserved event or a negative sample. Therefore, with the increase of sampled positive values $1$, their performances are constantly improved. 

\mypar{Effects of Positivity Biases in Matrix Completion.}
The curves of all three methods show a downward trend in the results of the positivity-oriented exposure with the density increasing. It is inevitable because the number of positive samples (i.e., ground-truth preferences) needed to predict decreases sharply with the density increases, and the remaining unexposed positive samples are actually loved by a few users and thus are hard to predict by the model. 
When the observed data has a larger probability of containing positive samples than the unobserved part, i.e., contains positivity bias, the task of completing all missing values in the matrix will be hard. The influence becomes more serious with positivity bias becoming more severe, which explains the downward trend of the curves.

\begin{figure}[!t]
  \tabcolsep=0pt
  \centering
  \includegraphics[width=.97\linewidth]{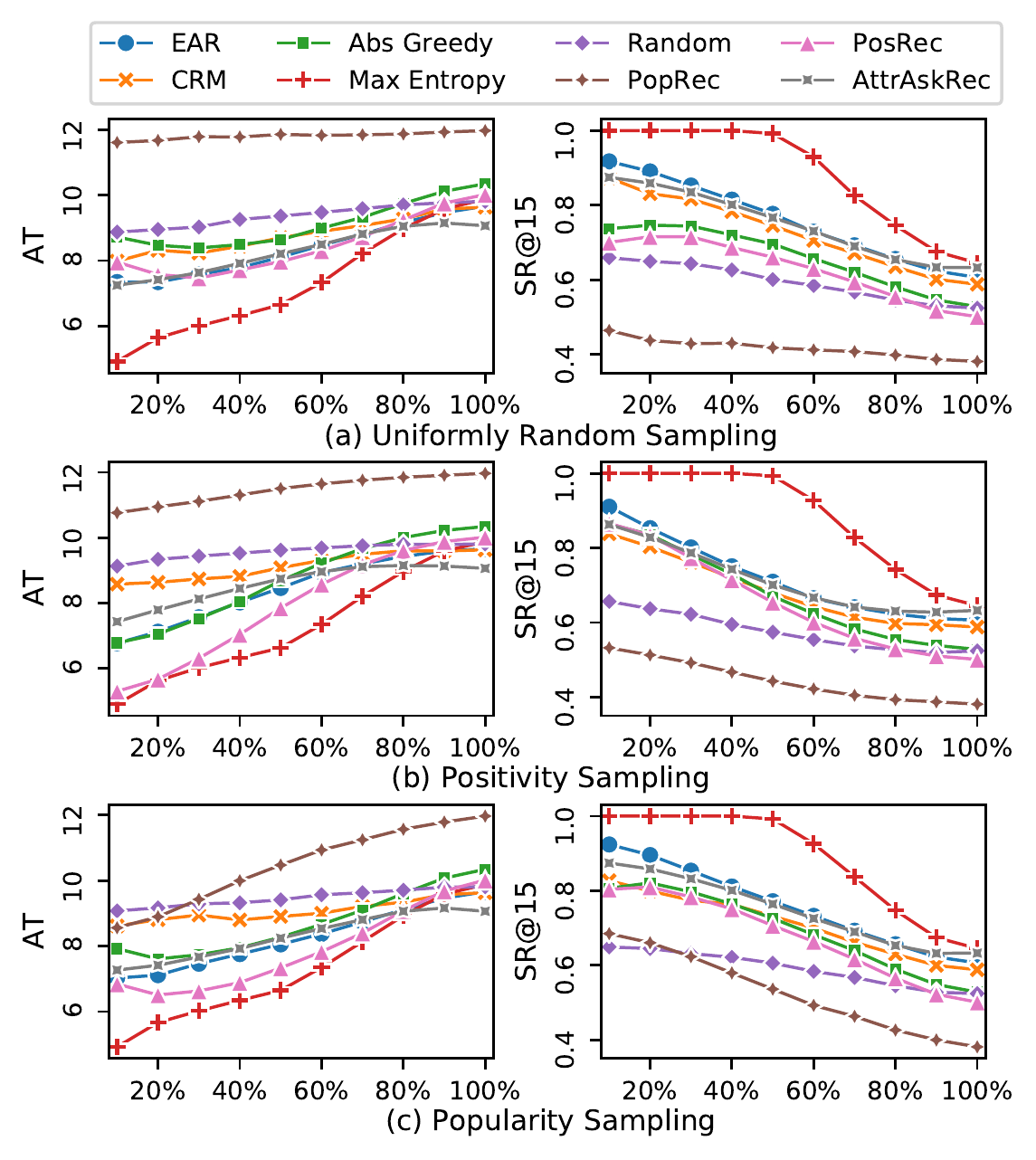}
   \vspace{-3mm}
  \caption {Performance of eight methods on the MTG setting using the estimated fully-observed data after completing the missing values via ExpoMF-cov.}
  \label{fig:MTG_estimated}
   \vspace{-4mm}
\end{figure}


\subsubsection{Analyses of The Effects of Estimated Values in Evaluation}
Due to the limited space, we only select ExpoMF-cov, which performs best in our experiments, to estimate the missing values in the partially-exposed data and evaluate all \acp{CRS} on the completed matrix. Note that the \acp{CRS} are pre-trained on the \emph{big matrix}.
The results of the CRS evaluation conducted on the completed data are illustrated in \myfig{MTG_estimated}. We calculate the number of the sampled datasets having inconsistent ranking of a CRS before and after matrix completion in \mytable{rankings}. 
An inconsistent ranking means the ranking of a CRS on the current test set is inconsistent with the counterpart on the fully observed data. For example, the ranking of \emph{EAR} is \emph{5} on the fully observed data and is \emph{7} on a sampled dataset under the positivity-oriented exposure when the density is $10\%$, then \emph{EAR} has an inconsistent ranking under the two different datasets.

\mypar{Effects of Completing the Partially-exposed Data.}
By analyzing the results w.r.t. SR@15 in \mytable{rankings}, we find that matrix completion can alleviate the inconsistency of \acp{CRS} in certain cases. For instance, under the random exposure, \emph{EAR} has inconsistent rankings on $7$ sampled partially-exposed datasets with different data densities. The number reduces to $5$ after estimating the missing values. There are nearly half of the values in the table decrease after matrix completion, and the improvement is evident under the random exposure.
This validates the helpfulness of matrix completion.

However, the matrix completion does not help restore the true rankings of \acp{CRS} w.r.t. AT, where only the number of \emph{PopRec} decreases.
We argue that it is because most \acp{CRS} have the indistinct absolute performance w.r.t. AT (see \myfig{MTG}), and the estimating error introduced by matrix completion can further blur the difference between them. By contrast, the absolute performances of different CRSs on SR@15 are more distinguishable, hence it is easier to restore the truth rankings after completing the matrix, even though the matrix completion introduces some errors.

\mypar{Effects of Biases after Estimating Missing Values.}
Generally, even after estimating the missing data, the number of inconsistent rankings of the uniformly randomly sampled data is smaller than the sampled data with biases. This indicates that the biases in the evaluation data persist to affect the CRS evaluation even after estimating the missing values. In some cases, completing the missing values can bring additional errors that make the evaluation worse.
Therefore, estimating the missing values in partially-exposed data can only help identify the true evaluation rankings of CRSs in certain cases where the biases are not serious. Therefore, these insights further highlight the importance of our fully-observed dataset.



\begin{table}[t!]
\caption{The number of the sampled datasets that have the inconsistent ranking of a CRS \emph{before/after} matrix completion.}
\renewcommand\arraystretch{1.2}
\begin{tabular}{@{}c|cc|cc|cc@{}}
\toprule
                     & \multicolumn{2}{c|}{\textbf{Random}} & \multicolumn{2}{c|}{\textbf{Popularity}} & \multicolumn{2}{c}{\textbf{Positivity}} \\
\textbf{Methods}     & \textbf{AT}     & \textbf{SR@15}     & \textbf{AT}       & \textbf{SR@15}       & \textbf{AT}       & \textbf{SR@15}      \\ \midrule
\textbf{EAR}         & 9/9             & 7/5                & 9/9               & 7/6                  & 9/9               & 6/5                 \\
\textbf{CRM}         & 9/9             & 5/0                & 9/9               & 4/4                  & 9/9               & 3/3                 \\
\textbf{Abs Greedy}  & 4/7             & 3/0                & 7/8               & 3/4                  & 6/7               & 3/3                 \\
\textbf{Max Entropy} & 5/9             & 0/0                & 9/9               & 0/0                  & 9/9               & 0/0                 \\
\textbf{Random}      & 5/9             & 3/8                & 9/9               & 9/8                  & 9/9               & 8/8                 \\
\textbf{PopRec}  & 1/0             & 1/0                & 3/2               & 3/2                  & 1/0               & 0/2                 \\
\textbf{PosRec}  & 5/9             & 2/8                & 9/9               & 8/8                  & 8/8               & 8/8                 \\
\textbf{AttrAskRec}   & 6/8             & 7/5                & 7/8               & 7/6                  & 7/8               & 6/5                 \\ \bottomrule
\end{tabular}
\label{tab:rankings}
\end{table}


\section{Conclusion}
In this paper, we present KuaiRec, a fully-observed dataset in recommender systems. We use this dataset to synthesize partially-exposed data to study how the data density and exposure bias in the traditional partially-observed data affect the evaluation of recommendations. Extensive experiments provide interesting insights into the evaluation of CRSs.

However, there are loose ends to our discussions. We hope our unique fully-observed data can support more research in a broader context. 
First, it can serve as a testbed for building trustworthy user simulators using partially-observed user-item interactions. Although the matrix completion in our experiments demonstrates limited help, it still remains an open question on whether it is possible to use partially-observed data to simulate fully-observed data correctly. Our fully-observed data can further support this exploration. 
Second, the fully-observed dataset can serve as a benchmark dataset for many research directions in recommendation, such as debiasing in recommender systems, interactive recommendation, and faithful evaluation. At last, after releasing this fully-observed data, we want to encourage the efforts of collecting more fully-observed datasets with richer properties, such as multiple domains or more diverse demographics.

\section{Acknowledgements}
This work is supported by the National Natural Science Foundation of China (61972372, U19A2079, 62102382).

\bibliographystyle{ACM-Reference-Format}
\bibliography{main}


\end{document}